\documentclass[journal]{new-aiaa}
\usepackage[english]{babel}
\usepackage[utf8]{inputenc}
\usepackage{textcomp}
\usepackage{graphicx}
\usepackage{amsmath}
\usepackage[version=4]{mhchem}
\usepackage{siunitx}
\usepackage{longtable,tabularx}
\usepackage{multirow}
\usepackage{dcolumn}
\usepackage{bm}
\usepackage{xcolor}

\title{Droplet shape representation using Fourier series and autoencoders}

\author{Mihir Durve \footnote{mihir.durve@iit.it}}
\affil{Center for Life Nano- \& Neuro-Science, Fondazione Istituto Italiano di Tecnologia (IIT), viale Regina Elena 295, 00161 Rome, Italy}

\author{Jean-Michel Tucny }
\affil{Dipartimento di Ingegneria, Università degli Studi Roma Tre, via Vito Volterra 62, Rome, 00146, Italy}
\affil{Center for Life Nano- \& Neuro-Science, Fondazione Istituto Italiano di Tecnologia (IIT), viale Regina Elena 295, 00161 Rome, Italy}

\author{Deepesh Bhamre}
\affil{Bhonsala Military College, Savitribai Phule
Pune University, Ram Bhoomi, Dr Moonje Marg, Nashik, India 422005}

\author{Adriano Tiribocchi}
\affil{Istituto per le Applicazioni del Calcolo del Consiglio Nazionale delle Ricerche, via dei Taurini 19, Roma, 00185, Italy}
\affil{INFN "Tor Vergata" Via della Ricerca Scientifica 1, 00133 Roma, Italy}

\author{Marco Lauricella}
\affil{Istituto per le Applicazioni del Calcolo del Consiglio Nazionale delle Ricerche, via dei Taurini 19, Roma, 00185, Italy}

\author{Andrea Montessori}
\affil{Department of Civil, Computer Science and Aeronautical Technologies Engineering, Roma Tre University, via Vito Volterra 62, Rome, 00146, Italy}

\author{Sauro Succi}
\affil{Center for Life Nano- \& Neuro-Science, Fondazione Istituto Italiano di Tecnologia (IIT), viale Regina Elena 295, 00161 Rome, Italy}
\affil{Department of Physics, Harvard University, 17 Oxford St, Cambridge, MA 02138, United States}

\begin{document}

\maketitle

\begin{abstract}
The shape of liquid droplets plays an important role in aerodynamic behavior and combustion dynamics of miniaturized propulsion systems such as microsatellites and small drones. Their precise manipulation can yield optimal efficiency in such systems. It is desired to have a minimal representation of droplet shapes using few parameters to automate shape manipulation using self-learning algorithms, such as reinforcement learning. In this paper, we use a neural compression algorithm to represent, with only two parameters, elliptical and bullet-shaped droplets represented with 200 points (400 real numbers) at the droplet boundary. The mapping of many to two points is achieved in two stages. Initially, a Fourier series is formulated to approximate the contour of the droplet. Subsequently, the coefficients of this Fourier series are condensed to lower dimensions utilizing a neural network with a bottleneck architecture. Finally, 5000 synthetically generated droplet shapes were used to train the neural network. With a two real numbers representation, the recovered droplet shapes had excellent overlap with the original ones, with a mean square error  $ \sim 10^{-3}$. Hence, this method compresses the droplet contour to merely two numerical parameters via a fully reversible process, a crucial feature for rendering learning algorithms computationally tractable.
\end{abstract}

\section{Introduction}

Droplet morphology is a critical determinant of performance and efficiency in combustion devices \cite{wang_2022}. Indeed, understanding droplet shape is pivotal for optimizing combustion processes, encompassing fuel consumption, emission reduction, and overall system performance \cite{FAETH,wang_2022, Momeni2013}. As astronautic and aeronautic devices operate under extreme conditions, droplet morphology influences key parameters such as heat transfer rates, spray dynamics, and pollutant formation. A combination of experimental techniques alongside numerical simulation methods have been used, for instance, to observe and generate detailed data on the characteristics of the combustion process
\cite{sirignano2014advances,boyd2024simulation,palmore2021vaporization,li2021heat,luo2021evaporation}, where it is of interest to control the shape of these droplets in real-time so as to maximize the quality of the 
process. Droplet shape also plays a key role in other applications, such as in drug delivery \cite{pais,pontrelli} and tissue engineering \cite{guzowski,tiribocchi,montessori,sh_au}. However, the process of droplet generation is inherently dynamic, presenting a challenge in effectively controlling not only droplet shape but also associated attributes, such as their volume, evaporation rate, and other physical observables.

Computer vision, a subset of machine learning algorithms, can be used to infer individual droplet boundaries from the 2D images captured from experiments or industrial processes. The desired shape of the droplets could be attained, for example, via automated feedback-based decision-making algorithms, such as reinforcement learning \cite{Sutton1998}, by controlling the droplet production apparatus in real time. As such, reinforcement learning could be used to control droplet shapes to optimize fuel burning 
and droplet production in the combustion process. However, reinforcement learning algorithms, such as Q-learning \cite{Watkins1992}, encounter computational challenges when the system's state is represented by a higher-dimensional vector.
Therefore, it is desirable to map the droplet's state to a lower-dimensional space to practically implement these algorithms. In this study, we achieve this transformation through a two-step process - firstly, by encoding the droplet's contour using Fourier descriptors, and secondly, by reducing the dimensionality of the Fourier descriptors through autoencoders.

Since the shape of a free liquid droplet is governed by the underlying physics, its boundary is free of pathological situations like kinks and discontinuities. This makes it possible to describe the contour of a liquid droplet via a smooth, well-behaved function, which in turn can be constructed from the coefficients of a Fourier series. In Refs.\cite{zhang2001comparative,gero},
for example, Fourier descriptors have been used to write a functional form of arbitrary closed shapes. 
They have also been useful in other applications, ranging from image analysis \cite{marques}, forensic anthropology \cite{Caple2017} and biometric identification using iris scan \cite{Rakshit} to the analysis of particle shapes \cite{schwarczFouriershape}, to name a few. In the context of droplets, their shape evolution is often studied by observing the change of Fourier descriptors of the droplet's contour \cite{kadivar}. In a similar spirit, in this study we derive a functional form of the contour delineating the boundary of a droplet by employing a Fourier series representation, whose descriptors are later mapped to lower dimensional vectors using autoencoder neural networks \cite{li_2023}.

The latter is an example of unsupervised learning technique that uses neural networks for representation learning. More specifically, this method implements a bottleneck in the network which maps the input data to a lower dimensional representation by learning the correlation between them.

By using a combination of Fourier series representation and autoencoder neural networks, this paper demonstrates a two-step fully reversible procedure to map a droplet's boundary shape to a vector consisting of only two real numbers. 

The remainder of this article is structured as follows: Section \nameref{section:Method} outlines the procedure employed to reduce the dimensionality of droplet shapes using Fourier descriptors and an autoencoder neural network. The performance of those neural networks is then shown in Section \nameref{section:Results}. Some final remarks will be given in Section \nameref{section:Conclusion}.

\section{Method}
\label{section:Method}

\begin{figure} [h]
\centering
\includegraphics[width=\linewidth]{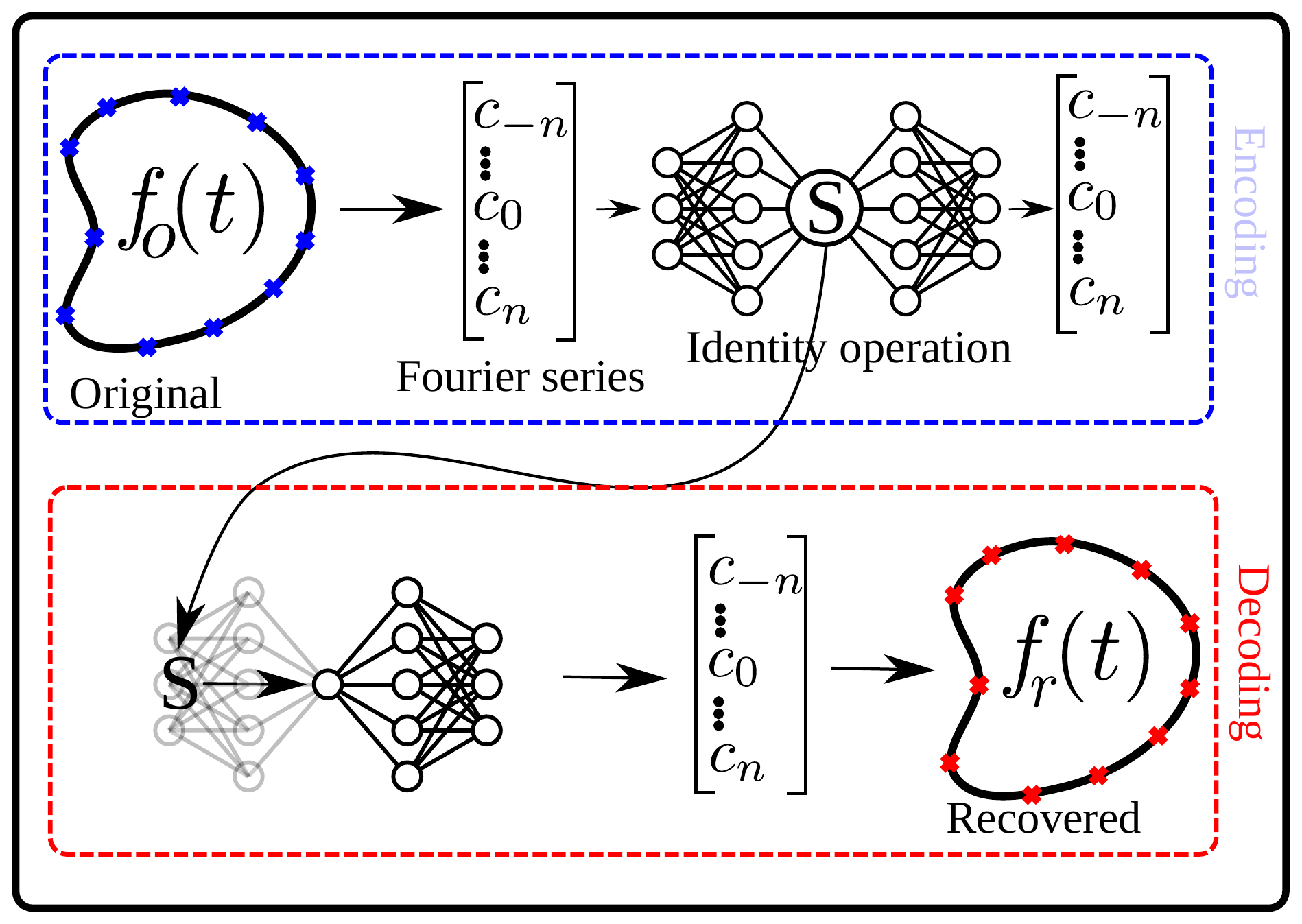}
\caption{Schematics of the method. During the encoding stage, the Fourier coefficients of the droplet contour are mapped to lower dimensional vector $S$ with a neural network having bottleneck architecture. During the decoding phase, starting from $S$, the Fourier coefficients are recovered using the decoder part of the neural network and later used to recover the droplet contour $f_r(t)$.  \label{Fig_method}}
\end{figure}

Figure \ref{Fig_method} illustrates the schematic depiction of a two-step procedure devised for compressing the droplet contour to a minimal representation. The droplet contour, i.e. the x,y coordinates of the interface of the droplet, can be extracted from 2D images through computer vision algorithms or through manual demarcation. This contour, denoted as $f(t)$ in a parametric form, is initially represented by a Fourier series formulation as $f_o(t) \approx f(t)$. Subsequently, the Fourier coefficients derived from this representation are employed in the training of a neural network featuring a bottleneck architecture. During the encoding phase, this neural network is structured to receive the Fourier coefficients of the original droplet contour $f_o(t)$ as input and is trained to predict the same coefficients at its output, effectively performing an identity operation. However, by doing so, the bottleneck layer of the trained network contains sufficient information for reconstructing the Fourier series coefficients. In instances where the number of nodes within the bottleneck layer is fewer than the count of Fourier modes, these ones are mapped onto a lower-dimensional vector denoted as $S$. Consequently, the entirety of information concerning the droplet contour is initially represented by a Fourier series, with the coefficients of this series further condensed into the lower-dimensional vector $S$.

The droplet contour can easily be recovered starting from the low dimensional vector $S$ with a reverse procedure. During this decoding phase, the vector $S$ is fed into the bottleneck layer as an input, prompting the network to generate Fourier series coefficients at its output. These coefficients are then used to reconstruct the shape $f_r(t)$. The methodology with operational details is described below.

\subsection{Fourier series description} 

We take a few representative droplet shapes generated by  lattice Boltzmann (LB) simulation and extract the liquid interface points forming a closed contour. Typically, we identify 200 equidistant points along the contour, which serve as input data for reconstructing the droplet shape given by $f(t)$. 
The x,y coordinates of these points are approximated by the function $f_o(t)$ with the interval of parameter $t$ being [0,1]. Since each droplet shape forms a closed contour, we identify the x,y coordinates at $t=0$ with those at $t=1$. In the complex form, the Fourier series representing $f_o(t)$ over this interval is given by:

\begin{equation}
 f_o(t) = \sum_{{n=-\infty}}^{n=+\infty} c_{n} e^{n \cdot 2\pi i t}.
 \label{eq_fourier}
\end{equation}

In practice, we terminate the series at some small integer value of $n$ which is enough to trace the contour of the droplet to a substantial precision. Here, the mean square error (MSE) between $f(t)$ and $f_o(t)$ is less than $10^{-4}$, essentially making $f_o(t) \approx f(t)$. 

The coefficients {$c_n$} of this Fourier series are given by:

\begin{equation}
 c_n = \int_{0}^{1} f_o(t) e^{-n \cdot 2 \pi i t} dt,
 \label{eq_coefficient}
\end{equation}

where $c_0 = \int_{0}^{1} f_o(t) dt$ represents the average of $f_o(t)$ and thus corresponds to the "midpoint" of the shape traced by $f_o(t)$.

These coefficients $c_n$ are then computed numerically by integrating Eq. \ref{eq_coefficient} as follows:

\begin{equation}
 c_n = \int_{0}^{1} f_o(t) e^{-n \cdot 2 \pi i t} dt \approx \sum_{t=0}^{t=1} [f_o(t) e^{-n \cdot 2 \pi i t} \Delta t].
\end{equation}

Here, $\Delta t = 1/k$, where $k$ represents the number of extracted data points. In this study, we have $k=200$ for each droplet contour. For a given $n$, the coefficients thus computed, viz. $c_{-n},...,c_{0},...,c_{n}$, are substituted back in Eq.\ref{eq_fourier} to reconstruct the droplet shape.

\begin{figure} [h]
\includegraphics[width=\linewidth]{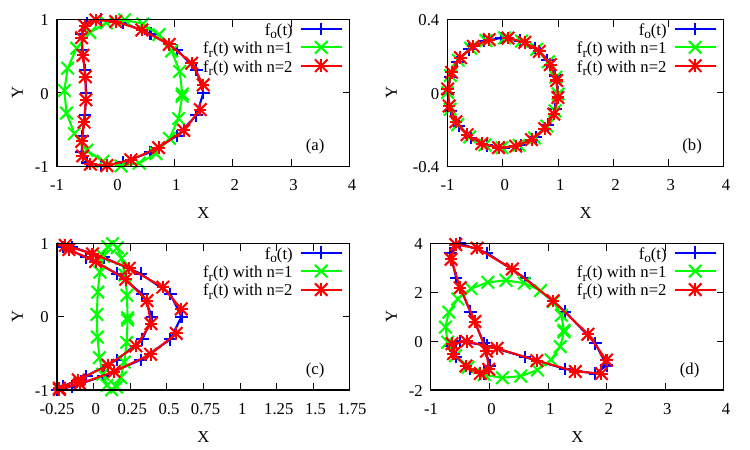}
\caption{Comparison of original $f_o(t)$ and regenerated droplet shapes $f_r(t)$ using Fourier series with various numbers of modes for a few selective shapes.}
\label{compare}
\end{figure}

In Fig.\ref{compare} (a)-(b), droplet shapes reconstructed in this manner are depicted for varying numbers of Fourier modes $n$ and compared with the original shapes extracted from the LB simulations. It may be noted that there is no need to normalize the droplet coordinates in order to write their functional form using Fourier descriptors. Computer vision algorithms often employ their own coordinate system for contour extraction and this feature offers an advantage to real-time decision-making algorithms as it eliminates the need for coordinate transformations. More specifically, we compare four contour shapes $f_o(t)$ and the regenerated contours $f_r(t)$ with increasing Fourier modes $n$.
(Note that, while the function $f_o(t)$ produced by Eq. \ref{eq_fourier} is inherently complex, we identify the imaginary axis with the real y-axis).

Out of such constructed contours, we limit our study to two fundamental morphologies discerned within LB simulations. The first morphology, termed "bullet-shaped" (Fig. \ref{compare} (a)), can be observed when a droplet is subjected to Poiseuille flow \cite{tiribocchi_PRE_2021} while the second one, comprising either circular or elongated droplets (Fig. \ref{compare} (b)), can be produced across diverse flow-centric devices \cite{ding_review2020, Davoodi, dangla, durve_2024}.
As expected, the reconstructed shape more closely aligns with the actual one as the number of Fourier modes increases. Importantly, our results show that two Fourier modes are sufficient for representing these basic shapes, and hence the contour of each droplet is fully captured by $2n+1=5$ complex numbers. 
Moreover, two Fourier modes have been found capable of generating a variety of other contours, such as the ones  shown in Fig. \ref{compare} (c)-(d), which in principle may not be relevant as feasible droplet shapes.

In the following step, a neural network  with bottleneck architecture 
will be trained, using the shapes described above, to further compress the five complex numbers (represented as ten real numbers in $x + iy$ format) to two real numbers. This will be done by performing an identity operation, thereby compressing the input to a lower dimensional vector.

\subsection{Training data compilation process} 
\begin{figure} [h]
\centering
\includegraphics[width=0.75\linewidth]{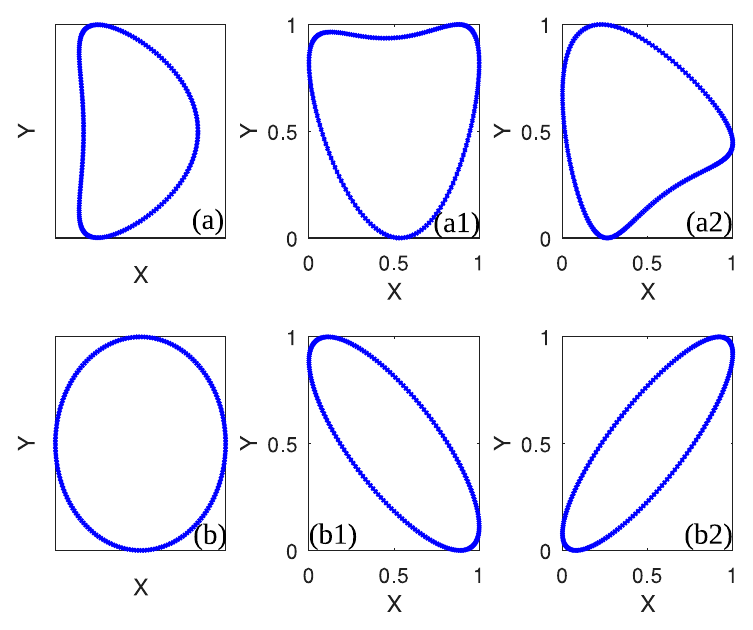}
\caption{Training data generated from two basic shapes. The two basic shapes, provided by LB simulations, are shown in the leftmost panels while the shape generated by adding noise and random rotation is shown by panels a1-a2, and b1-b2. \label{td}}
\end{figure}

For this study, we focus on two shapes shown in the leftmost panels of Fig.\ref{td} that are generated by LB simulations. The calculated coefficients for these shapes, denoted as $c_n$, offer a means to further construct other droplet contours. Initially, each coefficient $c_n$ is multiplied by uniformly distributed numbers sampled from $U[1-\eta,1+\eta]$. Here, we set the numerical value of $\eta=0.1$. In the second step, the resulting contour undergoes rotation by an angle $\theta$, randomly selected from the uniform distribution $U[-\pi,+\pi]$. The visual representation of this process is depicted in the central and right side panels of Fig. \ref{td}, which showcases four such generated contours. Notably, for this data generation, the Fourier series with $n=2$ was employed. Thus, the training data consists of  5000 1-D vectors, each vector consisting of 10 real numbers corresponding to the $2n+1$ complex coefficients of the Fourier series of individual droplet contours.

\subsection{Network architecture description} 
\label{subsection:Nad}
The neural network architecture comprises $10$ densely connected layers with varying numbers of neurons, starting with $50$ neurons in the first hidden layer, followed by hidden layers with $250$, $50$, $10$ neurons, outputting into the compressed information layer comprising only one neuron.  For each neuron of the hidden layers we use 
the parametrized rectified linear unit (PReLU) activation function \cite{he2015delving}, which differs from the classical ReLU activation function by a non-zero slope for negative input values. Preliminary trials indicate superior performance of the PReLU over ReLU and tanh activation functions, while the impact of varying numbers of neurons and layers was also explored in these trials, albeit without systematic presentation in this study. The total number of parameters for this neural network, calculated from the weights, biases, and hyperparameters associated with the PReLU activation function across all layers, amounts to $53168$. The neural network was trained on a medium-range laptop (intel i7 processor, 16 GB RAM) using a single core, taking about $30$ minutes to complete 50 epochs. The inference time was about $100$ milliseconds to recover the shape $f_r(t)$ starting from the state $S$ on the same computer. Thus, the training and inference exercises are computationally inexpensive and easily deployable.  

\section{Results}
\label{section:Results}
\subsection{Loss function during the training}

\begin{figure}[h]
\centering
\includegraphics[width=0.75\linewidth]{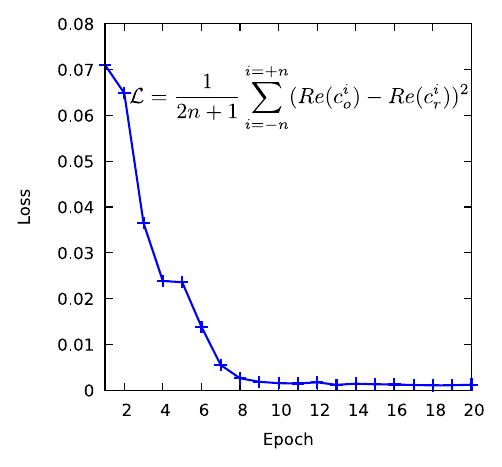}
\caption{Loss function for the entire data set as training progresses. The loss value ($\mathcal{L}$) is the mean square error between the five numbers at the input and five numbers at the output of a neural network. In the mathematical expression, $Re(c_o)$ and $Re(c_r)$ denote the real parts of the Fourier coefficients at the input and output of the network, respectively.\label{loss_value}}
\end{figure}

We conducted training of a deep neural network, as outlined in Section \ref{subsection:Nad}, utilizing a dataset comprising 5000 examples and 
employing custom code developed in TensorFlow.  
The quality of the training is quantified in terms of the loss function, which measures the difference between the predicted outputs and the actual target values.
Here, the loss function in use is the mean square error, which quantifies the difference between the input Fourier coefficients and the predicted ones at the output. In practice, we train a single neural network with the real part of the Fourier coefficients $Re(c_n)$ as input. Thus, the input and output of the neural network consists of five numbers. We recall that the network's objective is to execute identity operations, that is, to predict output identical to the input. 
As it can be seen in Fig. \ref{loss_value}, the loss function through the training procedure diminishes as training advances and ultimately reaches a plateau at a low value.

Subsequently, this trained neural network is employed to transform Fourier coefficients into a lower-dimensional vector $S$ in a two-step procedure. Initially, the real components of the coefficients (comprising five real numbers) are fed into the network, and the output at the bottleneck node, a singular value, is retained. Subsequently, the imaginary components of the coefficients are similarly processed, resulting in another value being stored from the bottleneck node. Consequently, each vector $S$ comprises of two real values, one derived from the compression of real components and the other from the imaginary components of the coefficients.

\subsection{Comparison of original and regenerated droplets} 
\begin{figure} [h]
\centering
\includegraphics[width=0.75\linewidth]{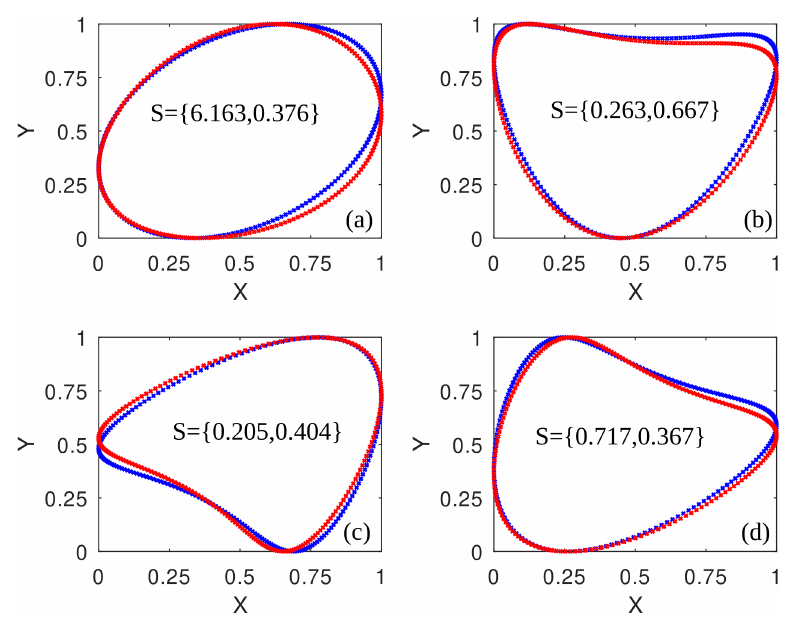}
\caption{Comparison between the original curve $f_o(t)$ in blue color and the recovered droplet contour $f_r(t)$ in red color. Vector $S$, showing the values at the bottleneck layer, is the compressed representation of the original contour. \label{compare2}}
\end{figure}

Upon training, the network is applied to a test dataset, which was not included in the training data. Figure \ref{compare2} displays the original function $f_o(t)$ (blue curve) and the regenerated function $f_r(t)$ (red curve) in four arbitrary instances from the test data. The input and output curves exhibit qualitative similarity.  

\subsection{Error distribution} 
\begin{figure} [h]
\centering
\includegraphics[width=0.75\linewidth]{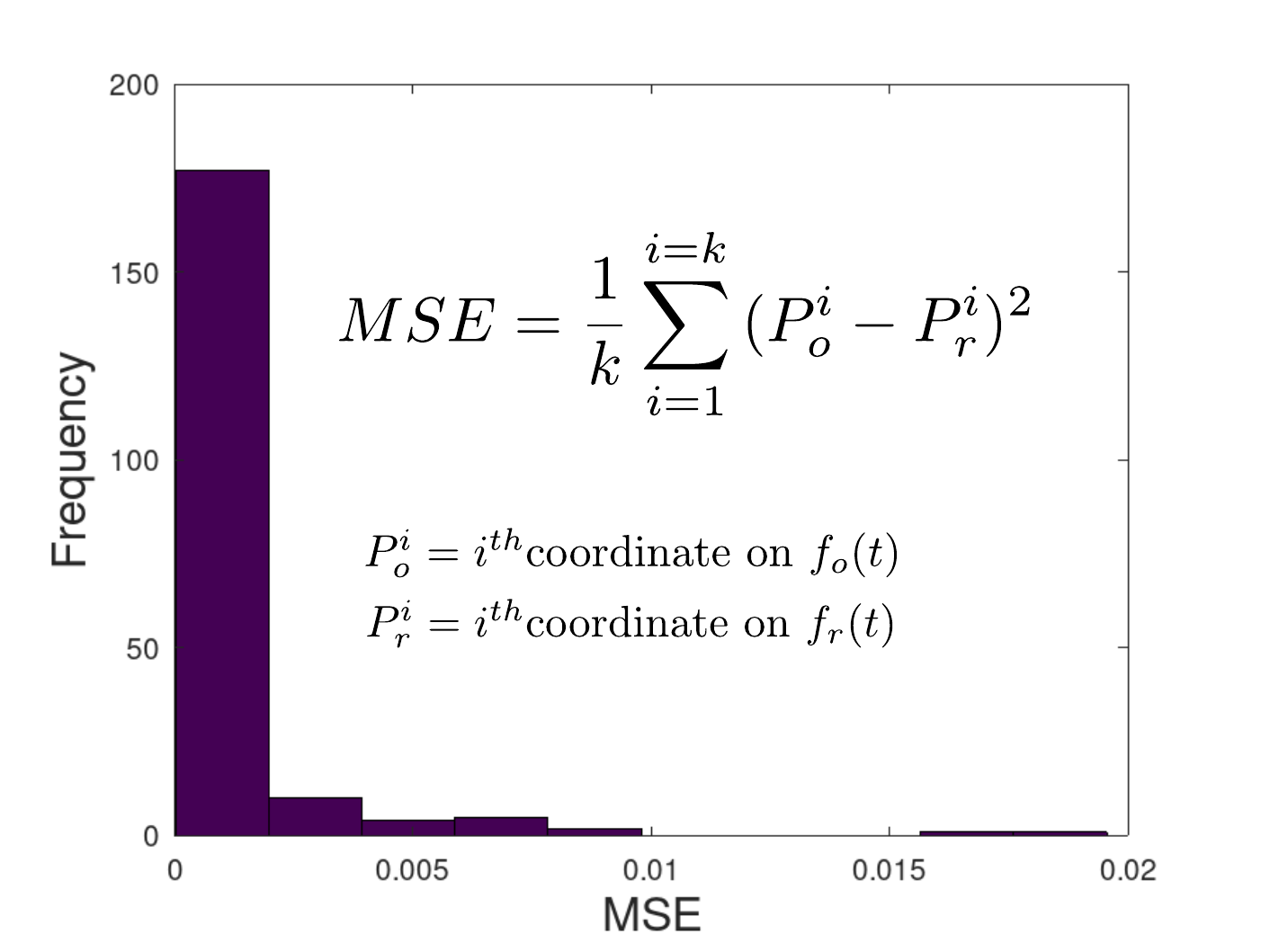}
\caption{Mean square error (MSE) distribution over 200 instances taken from the test data. The MSE is measured between the original droplet contour $f_o(t)$ and the regenerated droplet contour $f_r(t)$. Each regenerated droplet contour is recovered starting with a vector $S$ consisting of just two numbers. \label{distribution}}
\end{figure}

Figure \ref{distribution} illustrates the distribution of the mean square error (MSE) between the original contour $f_o(t)$ and regenerated contour $f_r(t)$ for 200 instances in the test data. The distribution's prominent peak is centered on 0.001, indicating an excellent correspondence between the two curves. Each contour is represented by $k=200$ coordinates.

\section{Conclusion}
\label{section:Conclusion}

Precise control of droplet shape is important to optimize the combustion process in astronautics and aeronautical devices. Although computational vision tools are able to generate large amounts of data for droplet shapes, there is a pressing need to represent those shapes using a small number of parameters, which can be used to automate control systems. In this paper, we aimed to find the minimal representation for the shape of droplets, using Fourier descriptors to preprocess the data and then neural networks to further compress the droplet shape representation.

It was first shown that five complex-valued Fourier descriptors (i.e a Fourier series with $n=2$) were sufficient to recover most of the observed droplet shapes. In this study, 5000 contours mainly derived from two typical shapes, namely ellipsoids and bullets, were used to train the neural network such that the trained fully-connected feedforward neural network was able to further reduce the representation to two real numbers. The mean square error between the original and recovered contours was observed to be the order of $10^{-3}$, thus efficiently compressing the original 400 data points to just 2 real numbers. 

Notably, the compressed contour representation can be used in reinforcement learning as a state ($s$) to learn optimal policy $\pi(a|s)$, i.e. to perform an action $a$ in a given state $s$, leading to a desired droplet shape in the combustion process. In other contexts, the lower dimensional droplet shape representation offers fewer parameters to study the shape evolution.  
Further studies will include an extension of the described methodology to more complicated shapes, such as the straining and breakage of droplets taken from experiments on dense emulsions and expanding the technique to represent droplets in 3D.

\section{Acknowledgment}
The authors acknowledge funding from the European Research Council ERC-PoC2 grant No. 101081171 (DropTrack). J.-M. T. thanks the FRQNT “Fonds de recherche du Québec – Nature et technologies (FRQNT)" for financial support (Research Scholarship No. 314328).  M.L. acknowledges the support of the Italian National Group for Mathematical Physics (GNFM-INdAM)

\bibliography{my_ref}

\end{document}